\documentclass[conference]{IEEEtran}
\IEEEoverridecommandlockouts
\usepackage{cite}
\usepackage{amsmath,amssymb,amsfonts}
\usepackage{url}
\usepackage{graphicx}
\usepackage{textcomp}
\usepackage{xcolor}
\usepackage{algpseudocode}
\usepackage{subfigure} 
\usepackage{algorithm2e}
\def\BibTeX{{\rm B\kern-.05em{\sc i\kern-.025em b}\kern-.08em
    T\kern-.1667em\lower.7ex\hbox{E}\kern-.125emX}}
\begin{document}

\title{Hybrid Neural/Traditional OFDM Receiver with Learnable Decider}

\author{Mohanad Obeed and Ming Jian\\
Huawei Technologies Canada Co., Ltd., Ottawa, Canada \\
\{mohanad.obeed, ming.jian\}@huawei.com}


%Telephone: (800) 555--1212, Fax: (888) 555--1212}


\maketitle

\begin{abstract}
%Deep learning (DL) based methods for orthogonal frequency division multiplexing (OFDM) radio receivers demonstrate higher signal detection performance compared to the traditional receivers. However, the dynamic nature of wireless channels may cause the DL models to fail or underperform in signal detection. This paper proposes a new architecture for OFDM receivers, where the neural and traditional receivers work together to improve the receiver robustness. 

Deep learning (DL) methods have emerged as promising solutions for enhancing receiver performance in wireless orthogonal frequency-division multiplexing (OFDM) systems, offering significant improvements over traditional estimation and detection techniques. However, DL-based receivers often face challenges such as poor generalization to unseen channel conditions and difficulty in effectively tracking rapid channel fluctuations. To address these limitations, this paper proposes a hybrid receiver architecture that integrates the strengths of both traditional and neural receivers. The core innovation is a discriminator neural network trained to dynamically select the optimal receiver whether it is the traditional or DL-based receiver according on the received OFDM block characteristics. This discriminator is trained using labeled pilot signals that encode the comparative performance of both receivers. By including anomalous channel scenarios in training, the proposed hybrid receiver achieves robust performance, effectively overcoming the generalization issues inherent in standalone DL approaches.

\end{abstract}


\section{Introduction}

The primary components of a communication system are the transmitter, receiver, and channel. Among these, the wireless channel is particularly challenging because it cannot be controlled or precisely predicted. To address this issue, transmitters and receivers are traditionally designed to continuously track channel variations and mitigate their effects. A common technique involves frequently sending pilot signals, enabling the receiver to estimate the channel and subsequently equalize and detect the transmitted symbols. Conventional methods typically utilize least squares (LS) estimation for channel estimation and linear minimum mean square error (LMMSE) for symbol equalization and detection. However, these methods struggle when the wireless channel undergoes rapid fluctuations, as frequent pilot signals alone become insufficient for accurate channel tracking.

\begin{figure}[t!]
\centering
\includegraphics[scale=0.35]{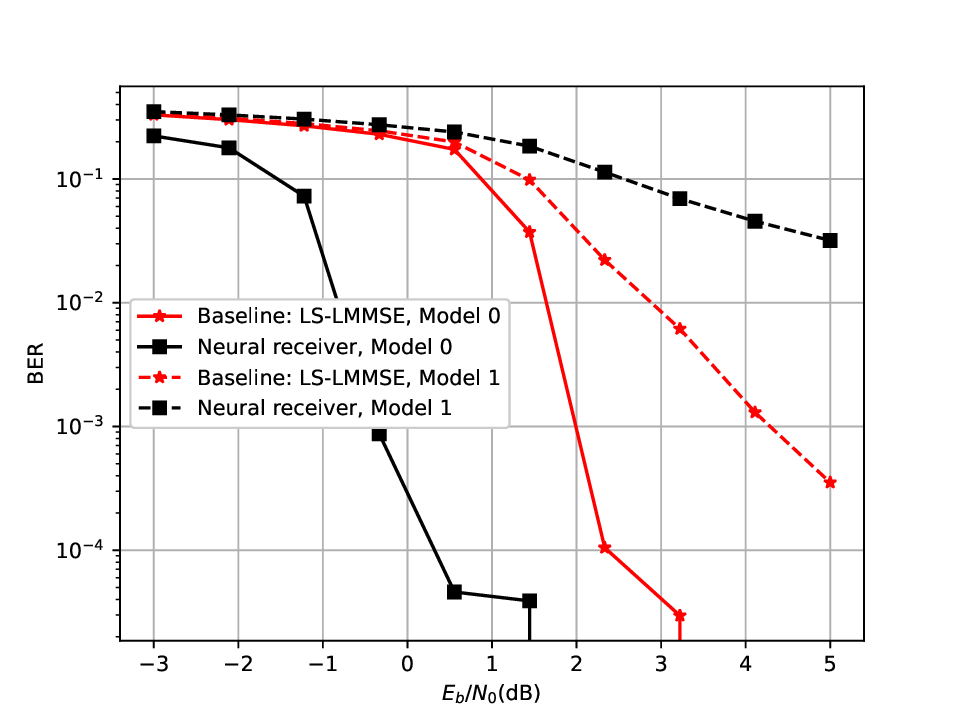}
\caption{Testing a neural receiver that is trained over channel model 0.  }
\label{fig:Motiv}
\end{figure}

To overcome the limitations of traditional approaches, deep learning (DL)-based receivers have been introduced, particularly for orthogonal frequency-division multiplexing (OFDM) systems, demonstrating significant performance enhancements over LS-LMMSE methods. Despite these improvements, DL models face challenges related to generalization when channel conditions deviate from those encountered during training. Specifically, DL receivers trained on one channel distribution may perform poorly under different or unseen channel conditions due to shifts in the actual channel distribution. For example, Fig. \ref{fig:Motiv} shows that a DL receiver trained on a clustered delay line (CDL) channel model (CDL-model C) with a delay spread of 100 ns performs effectively when tested on identical channel conditions. However, when evaluated on another CDL variant, such as CDL-model A with a delay spread of 1000 ns, its performance deteriorates, becoming worse than traditional approaches. Even with extensive training on diverse channel models, unique and previously unencountered channel characteristics can still lead to DL receiver underperformance. This variability suggests that while DL-based methods may outperform conventional approaches in certain scenarios, their performance can also be unpredictable and potentially inferior in dynamically changing channel environments.

Several papers have been proposed deep learning approaches to improve the receivers performance \cite{neumann2018learning, obeed2023alternating, chang2019complex, mei2022robust, 10571014, zhao2021deep, an2023learning, ait2021end, honkala2021deeprx}.  DL-based approaches either used to replace specific function in wireless receivers like the channel estimation or to replace multiple functions like channel estimation, equalization, and demapping. The authors of \cite{neumann2018learning} and \cite{obeed2023alternating} proposed to use neural networks for channel estimation. The authors of \cite{chang2019complex} utilized a convolutional neural network (CNN) for equalization, while the authors of \cite{mei2022robust} used a CNN as an equalizer and a demapper to calculate the log-likelihood ratios (LLRs). Those neural networks provided better performance compared to the traditional approaches used for channel estimation, equalization, or demapping. Other papers proposed to use neural networks to replace multiple functions in the receiver. The authors of \cite{10571014} used a CNN to jointly estimate the channels and detect the signals in single-carrier free-space optical receivers. The authors of \cite{zhao2021deep} used a CNN to directly detect the signals from the received OFDM time-domain signals. Other authors (e.g., \cite{an2023learning, ait2021end}) designed end-to-end deep learning based communication system, where the transmitter and the receiver are both trained jointly to form and detect the signals. The authors of \cite{honkala2021deeprx} designed a CNN with residual connections to handle the received frequency-domain OFDM signals and provide the soft bits as outputs. The designed receiver in \cite{honkala2021deeprx} (called DeepRx) utilized the frequency and temporal correlations in addition to the pilot signals in every received resource grid (RG) to improve the produced soft bits, which leads to minimizing the bit error rate (BER).

All the aforementioned works did not consider the dynamic nature of the wireless channels that may make the proposed deep learning models to fail due to the distribution shifts happened in the communication channels. The main concern of using neural networks is that they fail in some scenarios when the input distribution at the inference time deviates significantly from the input distribution used to train the models. To address the dynamic nature of wireless communications, researchers have proposed using adaptive deep learning models, such as meta-learning approaches, which can be fine-tuned online as the channel distribution shifts \cite{fischer2022adaptive, raviv2024adaptive}. The authors of  \cite{fischer2022adaptive, raviv2024adaptive} proposed adaptive neural networks at the receiver to track the channel distribution change. In general, the weights of the adaptive neural models must be tuned for every distribution shift to avoid providing disastrous results. However, these methods face several challenges, including the computational demands of online training, the necessity for true labels during each shift, and the difficulty in detecting distribution shifts early. These issues make it challenging to convince operators of the viability of adaptive AI approaches. The dynamic nature of the wireless channels, the chances of the traditional receivers to perform well in some scenarios, and the difficulty of employing adaptive learning motivate us to propose a new structure for wireless receivers.
%The invented receiver is based on employing both the deep learning and the traditional receivers according to the received signals distribution. 

This paper proposes a hybrid receiver that leverages the strengths of both traditional and neural receivers. Rather than entirely replacing the traditional receiver, we integrate a neural receiver to enhance overall system performance. This hybrid receiver dynamically selects the appropriate receiver based on the input data. Specifically, a small neural network, referred to as a discriminator, is trained to decide whether the received OFDM block aligns better with the neural receiver or the traditional one. The discriminator is trained using a dataset generated by comparing error rates from both receivers. Inputs consisting of received pilot signals are labeled based on whether the neural receiver or traditional receiver achieves fewer errors. This approach ensures that the discriminator captures the capabilities and limitations of both receivers. Additionally, incorporating anomalous channel conditions—unlikely but challenging scenarios—into the training dataset significantly enhances the discriminator’s ability to generalize across a wide range of channel environments. Numerical results show that the proposed hybrid receiver provides robust performance across diverse and dynamic channel conditions.

The rest of this paper is organized as follows. The considered system model is shown in Section II. Section III shows the proposed hybrid receiver and its different structures. In Section IV, we present the simulation results. Finally, our paper is concluded in Section V.

\section{System Model}

The proposed system model consists of traditional receiver, a DL-based receiver, and a learnable decider (discriminator). We assume an uplink channel, where the user-equipement (UE) is equiped with a single antenna and the receiver is equipped with multiple antennas $N$. However, extending this work to multiple-input multiple-output (MIMO) is straight forward. Initially, a sequence of uniformly distributed information bits is randomly generated. These bits undergo encoding using a low-density parity-check (LDPC) code then mapped to symbols, which are allocated across the available physical resource blocks within the transmission time interval (TTI). Demodulation pilots are inserted into designated subcarriers. Subsequently, the data is transformed into an OFDM waveform by applying an inverse fast Fourier transform (IFFT) to the residual blocks (RBs), producing 14 individual OFDM symbols per TTI. To mitigate inter-symbol interference (ISI), a cyclic prefix (CP) is appended to the beginning of each OFDM symbol before transmission.

We generated a total of 18 distinct channel models specified by 3GPP \cite{3gpp_tr38901_v16} for our analysis, comprising 6 clustered delay line (CDL) models, 6 tapped delay line (TDL) models, 2 urban microcell (UMi), 2 urban macrocell (UMa), and 2 rural macrocell (RMa) scenarios. The CDL and TDL models differ based on the specific channel types (A, B, and C) and their corresponding maximum delay spreads. Meanwhile, the UMi, UMa, and RMa models vary primarily in terms of outdoor-to-indoor penetration losses and shadow fading conditions, categorized into low and high-loss scenarios. For ease of reference, each of these generated channel models has been indexed sequentially from 0 to 17. The purpose of having diverse channel distributions is to evaluate the robustness of the proposed receiver and the ability to pick the best receiver for the given input signals.

Here, we explain the traditional and the DL-based receivers, respectively. Then in the following section, we explain how these receivers can be integerated with the help of the decider to provide better performance. 

\subsection{Traditional Receiver}

In a traditional receiver architecture, the received pilot signals within each RG are typically used to estimate the channel for every symbol in the grid. These estimated channels are then employed to equalize the received symbols. Following equalization, a demapper computes the log-likelihood ratios (LLRs) of the transmitted bits. In this context, we assume that the traditional receiver employs the least squares (LS) method for channel estimation, while the linear minimum mean square error (LMMSE) method is used for symbol equalization.

\begin{figure}[t!]
    \centering
    \subfigure[ResNet architecture]{
        \includegraphics[width=0.99\linewidth]{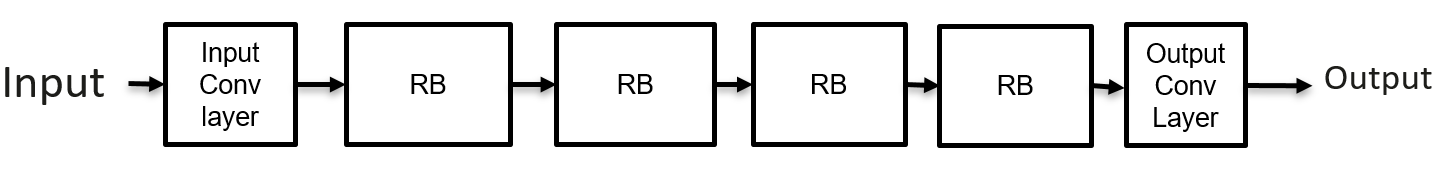}
        \label{fig:subfigResNet}
    }
    \hfill
    \subfigure[Residual block (RB) architecture]{
        \includegraphics[width=0.7\linewidth]{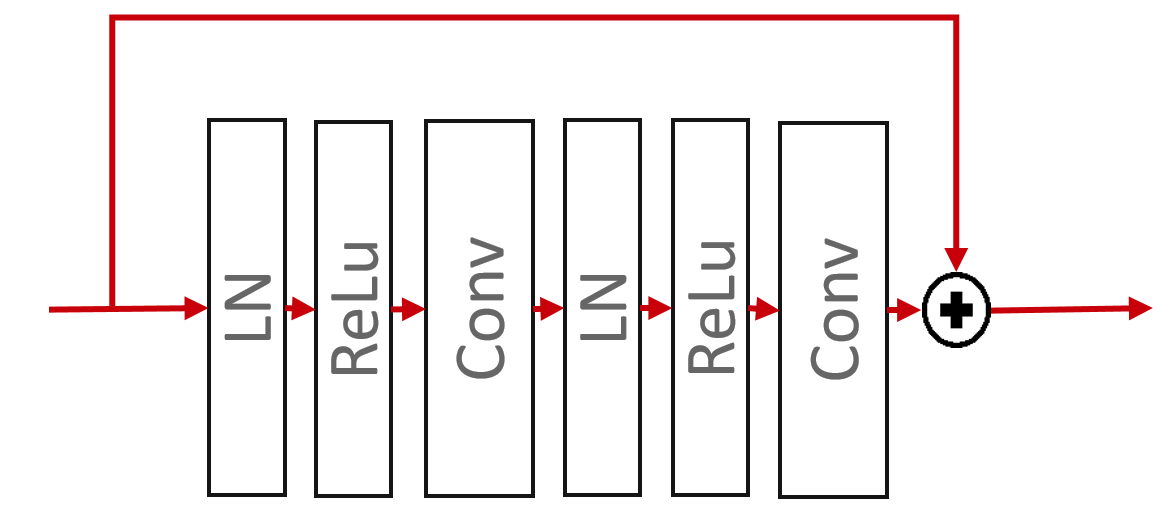}
        \label{fig:subfigRB}
    }
    \caption{The used ResNet in the hybrid receiver as a neural receiver. Conv means convolution layer,}
    \label{fig:resnet}
\end{figure}
\subsection{Deep Learning Based Receiver}
Various neural network architectures have been proposed to replace different receiver functions in communication systems. In this work, we adopt preactivation Residual Networks (ResNets) to jointly perform channel estimation, equalization, and demapping. The input to the ResNet consists of the received resource grids (RGs) concatenated with the known pilot symbols. We assume the ResNet architecture includes four RBs, as illustrated in Fig. \ref{fig:subfigResNet}, with the internal structure of each residual block detailed in Fig. \ref{fig:subfigRB}. Each convolutional layer within the network uses 128 filters with a kernel size of $3\times3$. The output of the network, optimized to represent log-likelihood ratios (LLRs), is reshaped appropriately to match the input format required by the channel decoder.

%\subsection{Generating and Organizing Datasets}

\section{The Hybrid Receiver}

 \begin{figure}[t!]
\centering
\includegraphics[scale=0.25]{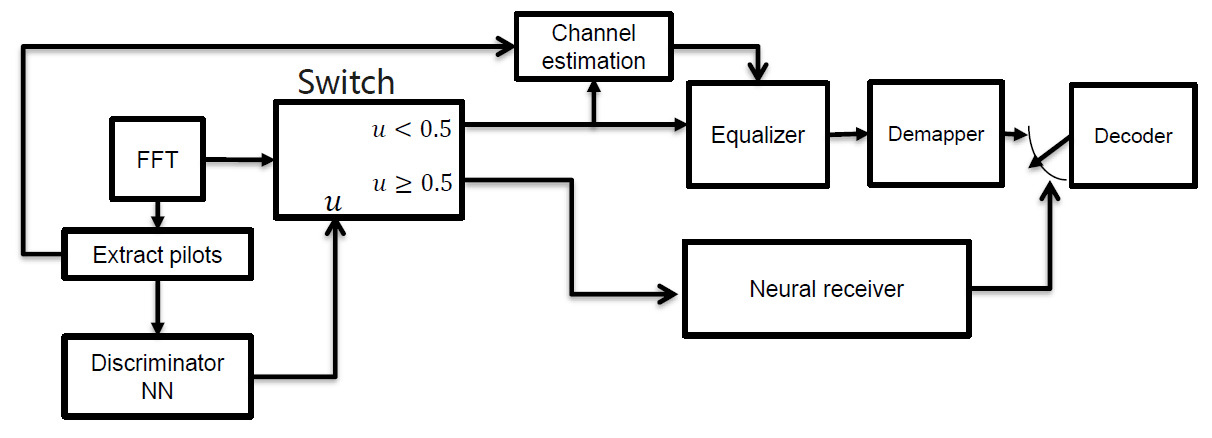}
\caption{Receiver Architecture I. If $u\geq 0.5$, the neural receiver will detect the LLRs using the FFT outputs signals; otherwise the traditional receiver will take over. }
\label{fig:Arch1}
\end{figure}
Traditional receivers rely on suboptimal methods for channel estimation, equalization, and demapping, often resulting in a large performance gap from the theoretical optimum. While deep learning (DL) approaches can help close this gap, their performance heavily depends on the training data distribution and often fails to generalize to unseen or dynamic wireless channel conditions, potentially leading to poor results. To address this, we propose a hybrid system that combines both DL-based and traditional receivers, with a learnable decider that selects the better option for each RG input. After training the DL receiver, a lightweight discriminator network is trained to identify whether a given input lies within the receiver's knowledge domain. We first describe the training procedures of both the DL receiver and the discriminator, followed by a discussion of the inference process.

During training, we first train the neural receiver using the provided channel dataset. Once trained, we extract the domain knowledge of both the neural and traditional receivers and embed it into a small neural network, referred to as the discriminator. To train this discriminator, we create a dataset where each input is a received pilot signal, and the label is 1 if the neural receiver produces fewer bit errors than the traditional receiver, and 0 otherwise. This allows the discriminator to learn which receiver performs better for each received OFDM block, effectively understanding the strengths and weaknesses of both receivers. Importantly, the training dataset includes some anomalous samples—rare channel conditions (e.g., extremely high delay spreads)—where the neural receiver is expected to underperform. Our results show that injecting samples from a single unknown channel distribution significantly improves the discriminator's generalization across a wide range of channel scenarios.

The main architecture of the proposed receiver is shown in Fig. \ref{fig:Arch1}, and it operates in the frequency domain. First, the time-domain received signals are transformed into frequency-domain signals using the fast Fourier transform (FFT). Assuming the signals are structured in an OFDM RG, each RG contains pilot symbols at known positions. These pilots are used as inputs to the discriminator neural network (NN), which outputs a value $u \in [0,1]$ using a sigmoid activation function. This value represents the probability of selecting the neural receiver for soft bit detection. A switch forwards the RG to the neural receiver if $u \geq 0.5$; otherwise, it routes it to the traditional receiver. The neural receiver reshapes the RG to match its input format and processes it to generate LLRs compatible with the decoder input. The decoder connects to the neural receiver output if 
$u \geq 0.5$, or to the traditional demapper otherwise. Both the neural receiver and the discriminator are trainable networks, but the discriminator is much smaller in size.

The discriminator is a CNN composed of three convolutional layers followed by ReLU activations and batch normalization, a flatten layer, one dense layer with 30 units, and a sigmoid output layer. The convolutional layers use 32, 64, and 64 filters, respectively. The total number of parameters in the discriminator is approximately 300K, which is small compared to the 1.2M parameters of the neural receiver.

\subsection{Different Hybrid Architectures}
 \begin{figure}[t!]
\centering
\includegraphics[scale=0.24]{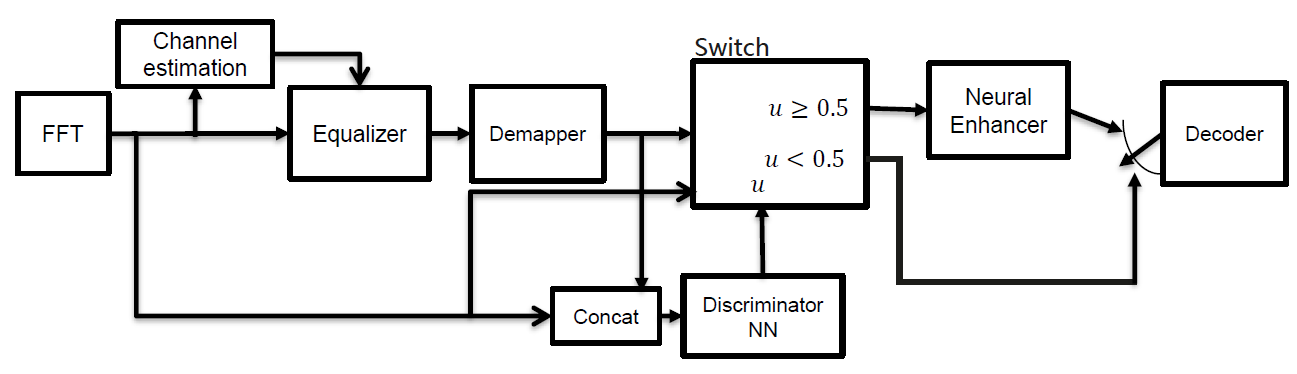}
\caption{Receiver Architecture II. If $u<0.5$, the switch forward the demapper's outputs (LLRs) to the decoder; otherwise, the LLRs will be concatenated with the FFT outputs and forwarded to the neural enhancer.  }
\label{fig:Arch2}
\end{figure}

 \begin{figure}[t!]
\centering
\includegraphics[scale=0.25]{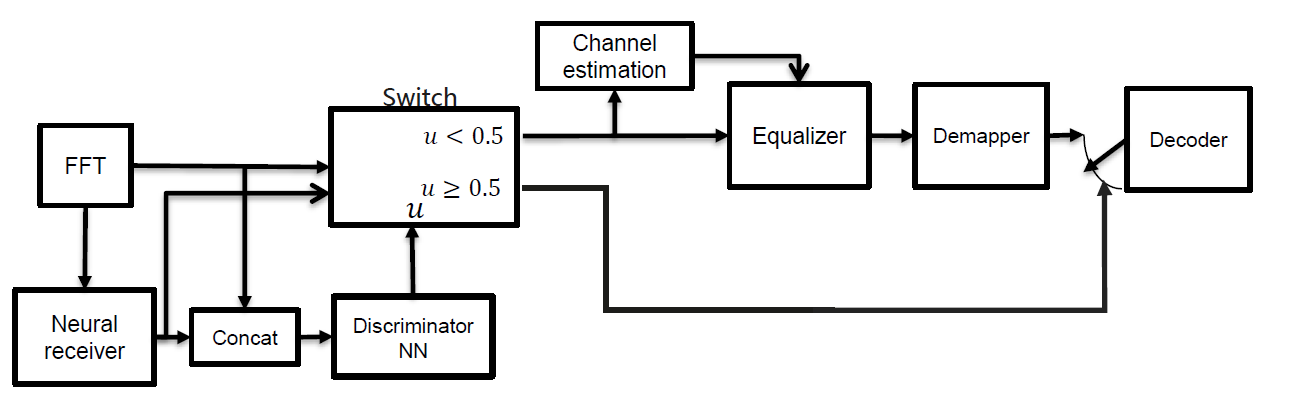}
\caption{Receiver Architecture III. If $u\geq 0.5$, the switch forward the neural's outputs (LLRs) to the decoder; otherwise, the switch forwards the FFT outputs the traditional receiver.}
\label{fig:Arch3}
\end{figure}
The hybrid receiver can be designed in different architectures. For instance, the received signals can be first processed by the traditional receiver. Then a discriminator NN can be used to decide whether the processed signal is required to be enhanced by a neural network or not. This is illustrated in Fig. \ref{fig:Arch2}. In this figure, the received signals are first handled by the traditional receiver to find LLRs. These LLRs concatenated with the input signals can be fed into a discriminator to decide whether these LLRs are sufficiently good or a neural network can be used to enhance it more. If the output of the discriminator is less than 0.5, the switch forwards the resulted LLRs from the traditional receiver into the decoder. Otherwise, the LLRs with the input signals will be forwarded to the NN enhancer to produce better LLRs.  

Another architecture of the hybrid receiver can be designed to first handle the received signals  by a neural receiver. Then based on the neural receiver outputs, the discriminator will decide whether the resulted LLRs are good or not with the given input signals. If the discriminator decides that the LLRs are good enough, it will produce a value of $u$ that is greater than 0.5, telling the switch to pass the LLRs to the decoder. Otherwise, (i.e., $u < 0.5$), the received signals (FFT signals) will be passed by the switch to the traditional receiver that will produce the LLRs to the decoder.  

In general, in the architecture shown in Fig \ref{fig:Arch1}, the paths are selected at the beginning via the discriminator and either the neural receiver or the traditional one will handle the LLRs production. In Fig. \ref{fig:Arch2}, the received signal must be processed first by the traditional receiver and the discriminator will decide whether the neural receiver is needed or not. In contrast, in Fig. \ref{fig:Arch3}, the received signal must be first handled by the neural receiver and then the discriminator will decide to use the traditional receiver or not.    

The main advantage of Architecture I is its low computational complexity. This is because the discriminator operates only on pilot signals, which have significantly lower dimensionality compared to the entire received RG, and only one receiver is activated per input RG. In contrast, Architectures II and III are more computationally demanding but offer higher accuracy in selecting the optimal receiver for each input. This improved accuracy comes from the discriminator’s use of both the full input RG and the log-likelihood ratio (LLR) outputs from the first receiver, enabling more informed decisions. However, this approach increases computational cost, as the discriminator must be larger to handle the high-dimensional input, and one or both receivers may be engaged to produce the final LLR outputs.

Therefore, due to space limitations, our simulation results focus solely on Architecture I. If this architecture demonstrates strong performance, it can be reasonably expected that Architectures II and III—designed for higher accuracy—will perform even better. Moreover, Architecture I is particularly attractive due to its low computational complexity, making it a practical choice for resource-constrained scenarios.

\begin{figure}[t!]
    \centering
    \subfigure[Level I training with 4-QAM]{
        \includegraphics[width=0.77\linewidth]{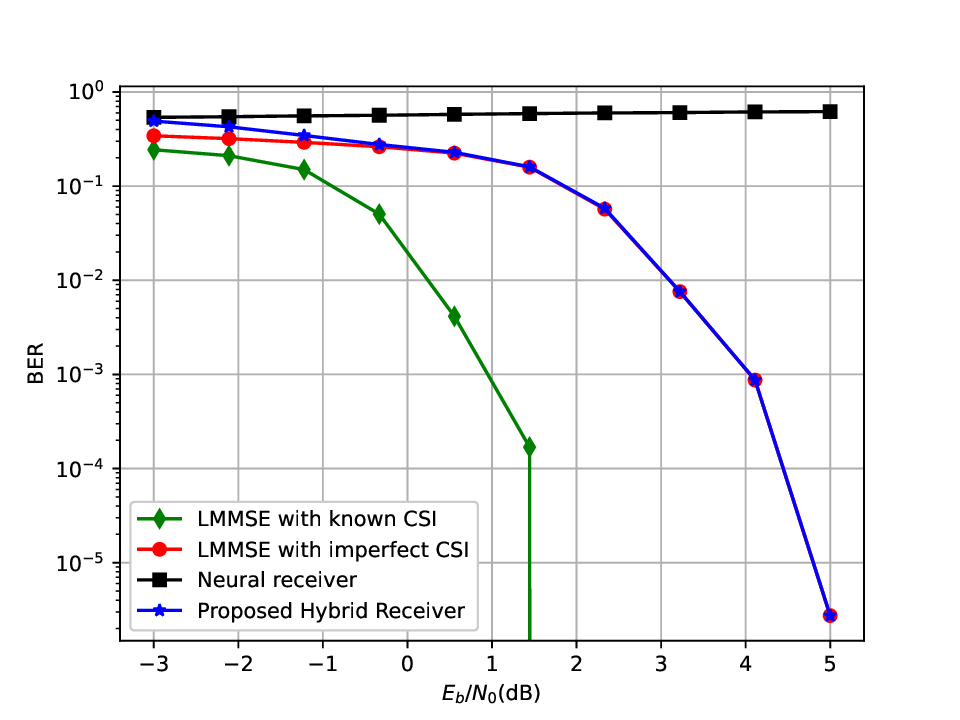}
        \label{fig:subfig42}
    }
    \hfill
    \subfigure[Level II training with 64-QAM]{
        \includegraphics[width=0.77\linewidth]{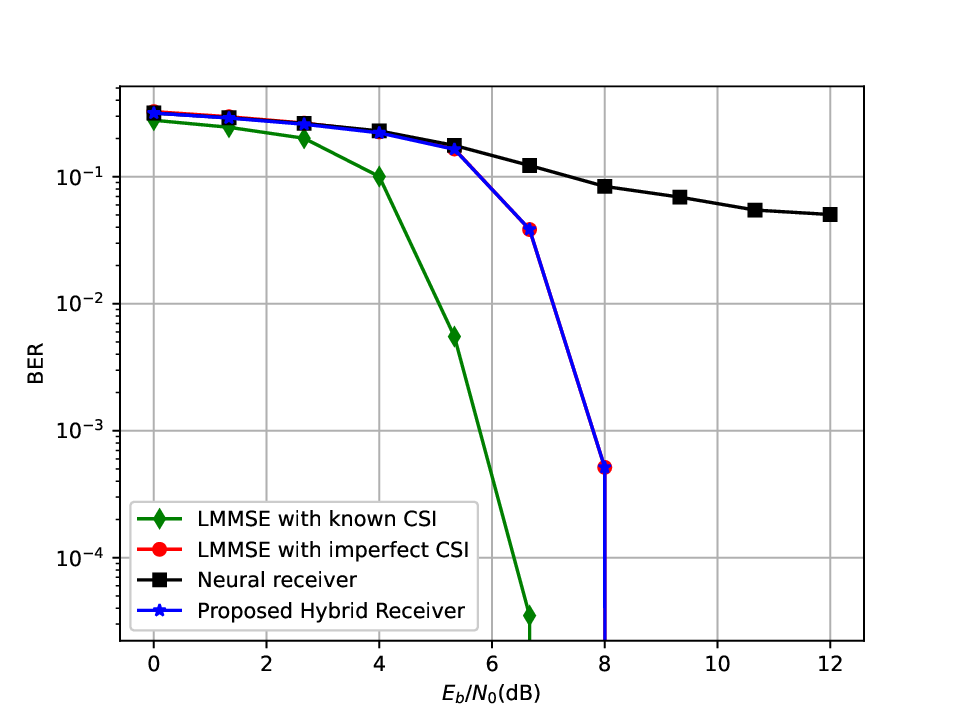}
        \label{fig:subfig642}
    }
    \caption{Comparison of receiver performance across different training levels and modulation schemes, when the index of the channel distribution (CI) equals to 2.}
    \label{fig:combined2}
\end{figure}

\section{Simulation Results}

In our simulations, we generated 18 distinct channel models, including 6 CDL, 6 TDL, and 2 each of UMi, UMa, and RMa. The CDL and TDL models differ by type (A, B, C) and maximum delay spread values. For the UMi, UMa, and RMa models, two versions were created: one with low outdoor-to-indoor loss and no shadow fading, and another with high loss and shadow fading. These models are indexed from 0 to 17. Sionna, a Python-based library, has been used to simulate the channels \cite{hoydis2022sionna}. We evaluate the performance of the proposed hybrid receiver, a standalone neural receiver, an LMMSE receiver with imperfect CSI, and an LMMSE receiver with perfect CSI. To explore the generalization ability of the DL-based receivers, we consider two training levels: 
 Level I training, where the neural receiver is trained on channel indices 0, 1, and 7, also using anomalies from index 4; Level II training, where the neural receiver is trained on the indices 0, 1, 3, 5, 7, 9, and 13, with anomalies also generated from index 4. In general, the figures evaluate the receivers with 4 quadrature amplitude modulation (4-QAM) and 64-QAM modulation schemes. We used Level I training to evaluate 4-QAM modulation, while Level II training is used to evaluate 64-QAM modulation. However, swapping the training levels across modulation schemes yields similar trends and conclusions.

% \begin{figure}[t!]
% \centering
% \includegraphics[scale=0.53]{Level3_CI2_64QAM.eps}
% \caption{..}
% \label{fig:SISO-1bt}
% \end{figure}

% \begin{figure}[t!]
% \centering
% \includegraphics[scale=0.53]{Level2_CI2_4QAM.eps}
% \caption{..}
% \label{fig:SISO-1bt}
% \end{figure}

\begin{figure}[t!]
    \centering
    \subfigure[Level I training with 4-QAM]{
        \includegraphics[width=0.77\linewidth]{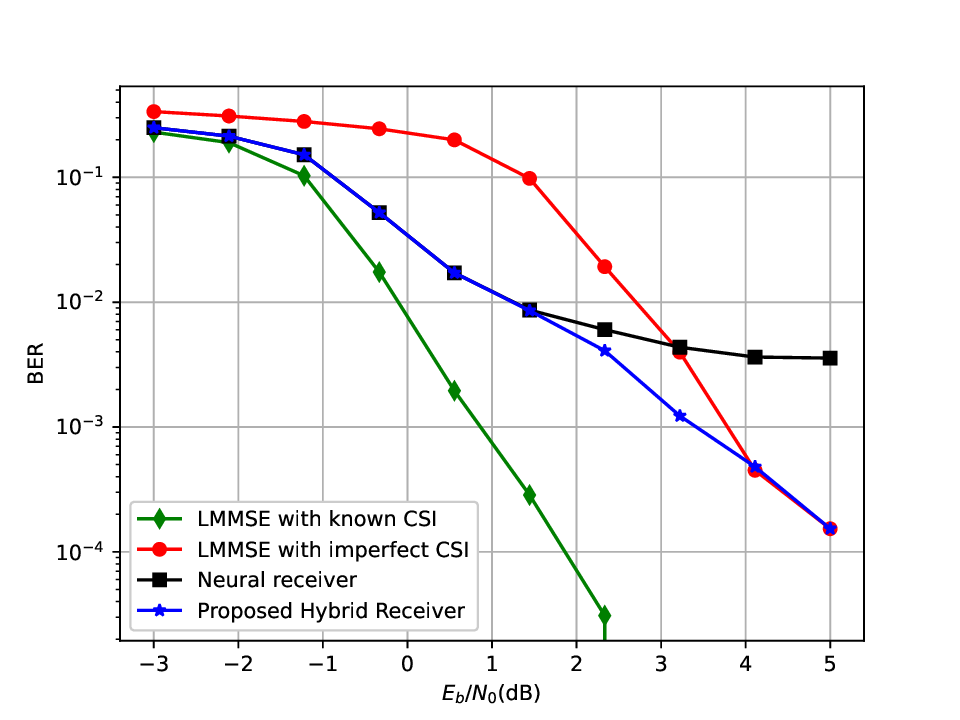}
        \label{fig:subfig412}
    }
    \hfill
    \subfigure[Level II training with 64-QAM]{
        \includegraphics[width=0.77\linewidth]{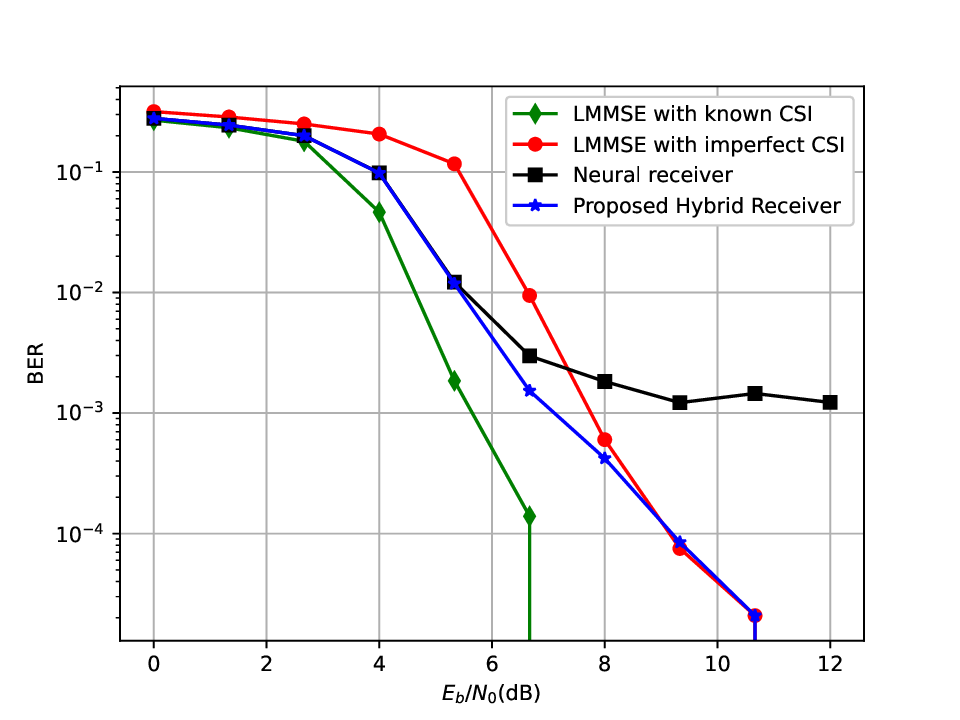}
        \label{fig:subfig6412}
    }
    \caption{Comparison of receiver performance across different training levels and modulation schemes, when the index of the channel distribution (CI) equals to 12.}
    \label{fig:combined12}
\end{figure}
We evaluate all receivers across the 18 considered channel distributions (channel indices, CIs), and results show that the discriminator accurately selects the better receiver for each input with high precision. However, due to space limitations, we present results for selected CIs that highlight key behaviors—such as cases where the neural receiver outperforms, where the LMMSE receiver performs better, where both perform similarly, and overall average trends.  For instance, Fig. \ref{fig:combined2} shows how the receivers behave when the channels are generated with the distribution of CI=2. From both figures \ref{fig:subfig42} and \ref{fig:subfig642}, the neural receiver fails to generalize for this kind of channel distribution whether the modulation scheme is 4-QAM or 64-QAM. This means that this channel distribution is completely different from those channels' distributions used to train the neural receiver. The figures also show that the proposed hybrid receiver with high accuracy selects the traditional receiver to operate for the received signals.

\begin{figure}[t!]
    \centering
    \subfigure[Level I training with 4-QAM]{
        \includegraphics[width=0.77\linewidth]{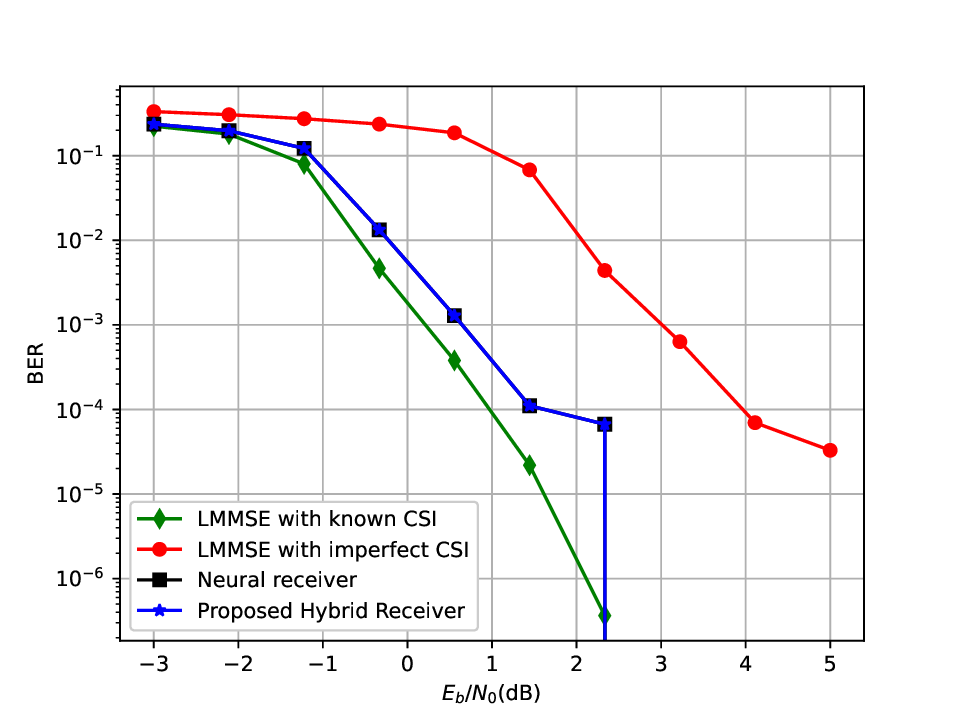}
        \label{fig:subfig46}
    }
    \hfill
    \subfigure[Level II training with 64-QAM]{
        \includegraphics[width=0.77\linewidth]{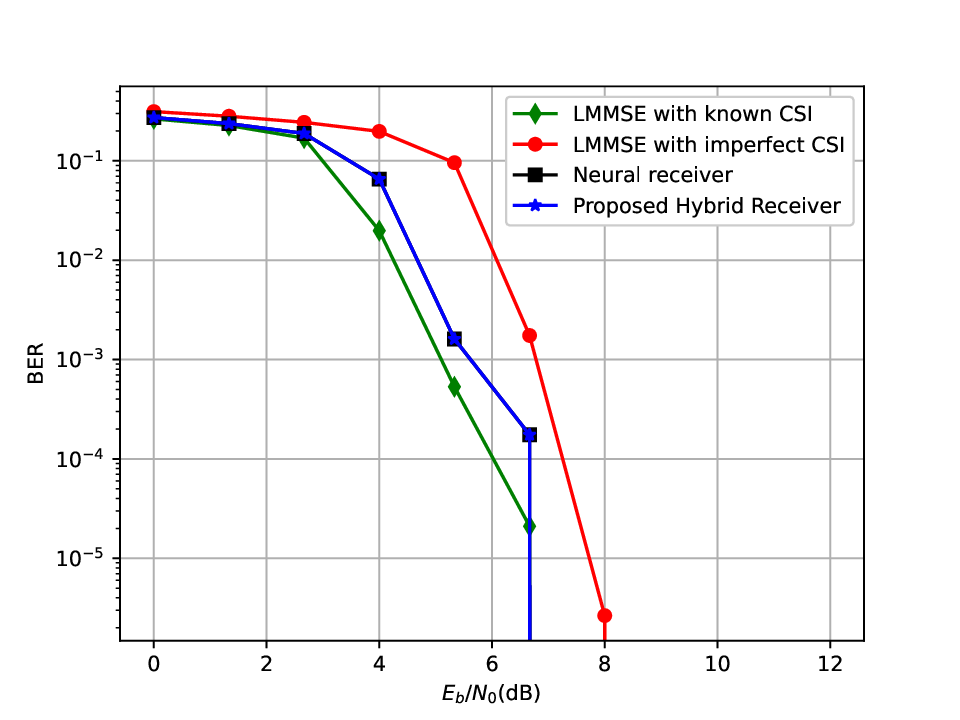}
        \label{fig:subfig646}
    }
    \caption{Comparison of receiver performance across different training levels and modulation schemes, when the index of the channel distribution (CI) equals to 6.}
    \label{fig:combined6}
\end{figure}

%  \begin{figure}[t!]
% \centering
% \includegraphics[scale=0.53]{Level3_CI12_64QAM.eps}
% \caption{..}
% \label{fig:SISO-2bt}
% \end{figure}
An illustrative case where the neural receiver performs well at certain SNR levels but underperforms at others is CI = 12. As shown in Fig. \ref{fig:combined12}, the neural receiver outperforms at low SNRs but performs poorly at high SNRs for both 4-QAM and 64-QAM modulations. In this scenario, the hybrid receiver effectively adapts by selecting the neural receiver at low SNR and switching to the traditional receiver at high SNR. Notably, the discriminator occasionally outperforms both individual receivers by dynamically choosing the better one for each input, thereby minimizing the overall BER.

% An example of distribution that shows that the neural receiver performs better with a range of SNR values and underperform at other range is when we consider CI = 12.  Fig. \ref{fig:combined12} shows that the neural receiver performs well with low SNR and poor when the SNR is high with 4 and 64 QAM modulations. It can be seen that the hybrid receiver in such a case selects the neural receiver for the low SNR and the traditional receiver for the high SNR. The discriminator at some points shows better performance than both receivers due to the fact that it alternates between both receivers at every input signal to minimize the BER. 

A strong example demonstrating the effectiveness of the neural receiver is CI = 6. The neural receiver generalizes well to this channel because its distribution closely resembles those seen during training. Fig. \ref{fig:combined6} illustrates the behavior of the hybrid receiver when the neural receiver performs effectively. As shown in both Fig. \ref{fig:subfig46} and Fig. \ref{fig:subfig646}, the discriminator consistently selects the neural receiver across all SNR values.

% A good example that shows the excellence of the neural receiver is to test the receivers on the CI=6. The reason that the neural receiver generalize for this channel index is that its distribution is quite similar to that distribution used for training the model.  Fig. \ref{fig:combined6} shows how the hybrid receiver will behave when the neural receiver performs well. Both figures (\ref{fig:subfig46} and \ref{fig:subfig646}) ahow that the discriminator select the neural receiver for all the SNR values. 

% \begin{figure}[t!]
% \centering
% \includegraphics[scale=0.53]{Level3_CI8_64QAM.eps}
% \caption{..}
% \label{fig:SIMO-1bt}
% \end{figure}

\begin{figure}[h!]
    \centering
    \subfigure[Level I training with 4-QAM]{
        \includegraphics[width=0.77\linewidth]{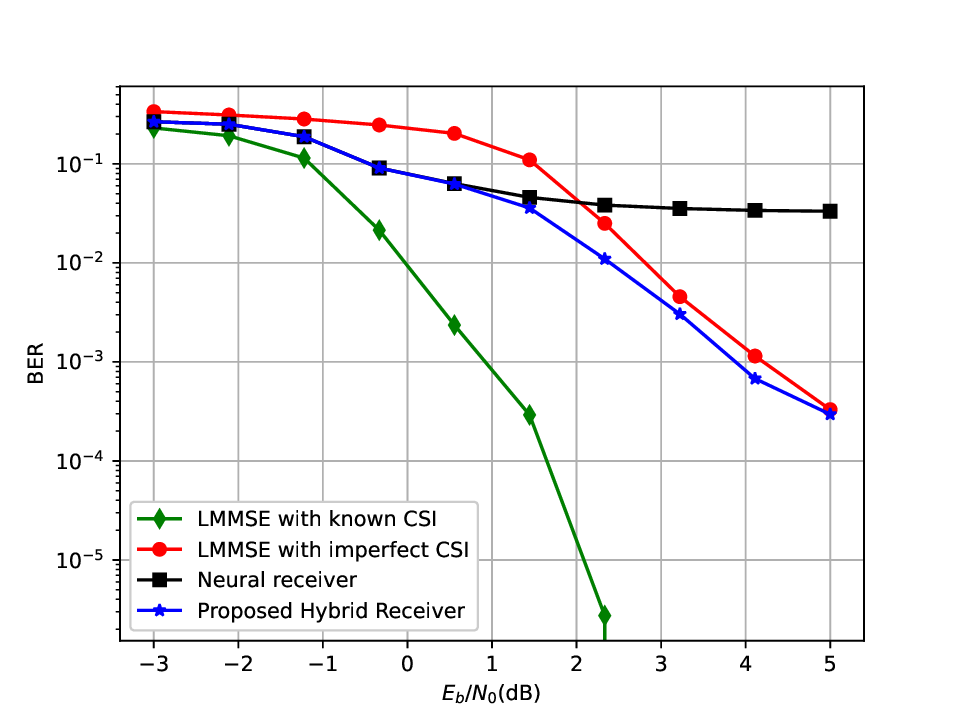}
        \label{fig:subfig4all}
    }
    \hfill
    \subfigure[Level II training with 64-QAM]{
        \includegraphics[width=0.77\linewidth]{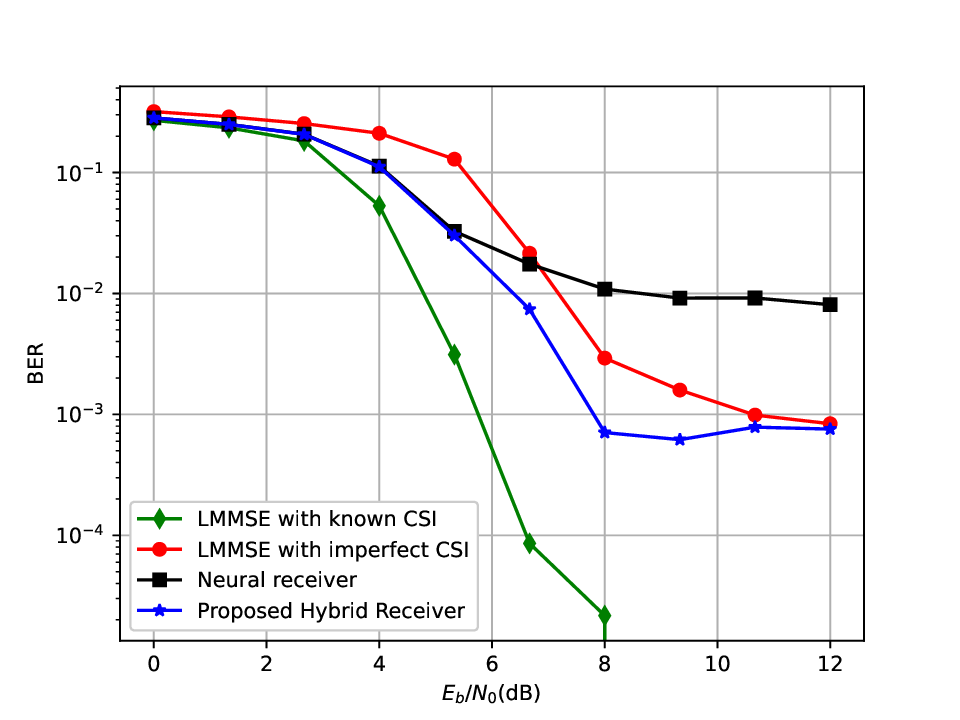}
        \label{fig:subfig64all}
    }
    \caption{Comparison of receiver performance across different training levels and modulation schemes, when the index of the channel distribution is randomly selected between 0 and 17.}
    \label{fig:combinedAll}
\end{figure}
To evaluate the overall performance of the proposed receiver, we test the system under random channel conditions, where each transmitted RG experiences a channel drawn from a random index between 0 and 17. Fig. \ref{fig:combinedAll} shows that the hybrid receiver either matches the performance of the better receiver or outperforms both. This is because the hybrid approach dynamically switches between the neural and traditional receivers to minimize the BER.

% To obtain the general picture of the proposed receiver, we examine the receivers when the channel distribution is randomly selected from the overall considered indices (0-17) for each input. In other words, for each transmitted RG, the channel is generated from a random index between 0 and 17.  Fig. \ref{fig:combinedAll} shows that the hybrid receiver either match the best receiver performance or performs better than both of them. This is because of that the proposed hybrid receiver alternates between both receivers to minimize the BER.

Across all figures, the proposed hybrid receiver effectively prevents the neural receiver from performing poorly under significant distribution shifts, while leveraging its strengths to achieve substantial BER reduction.

% From all of the above figures, the proposed hybrid receiver prevents the neural receiver from providing disastrous performance when there is a significant channel distribution shift and leverages the neural receivers to significantly minimize the BER. 
% \begin{figure}[t!]
% \centering
% \includegraphics[scale=0.53]{Level3_Overall_64QAM.eps}
% \caption{..}
% \label{fig:SIMO-2bt}
% \end{figure}

\section{Conclusion}
This paper proposed a hybrid OFDM receiver that combines neural and traditional (LMMSE-based) receivers to leverage the strengths of both while avoiding their individual weaknesses in dynamic wireless channel environments. The key component of the system is a learnable discriminator that evaluates each input signal and selects the appropriate receiver based on its learned understanding of each receiver's capabilities. Simulation results demonstrate that the discriminator generalizes better than the neural receiver, requiring only one anomalous distribution to effectively detect when the neural receiver may fail. Overall, the discriminator consistently selects the better receiver with high accuracy. Additionally, we explored different integration schemes between the neural and LMMSE receivers, highlighting trade-offs between performance gains and computational complexity.

\bibliographystyle{ieeetr}
\bibliography{Ref.bib}

\end{document}